\documentclass[12pt]{article}

\usepackage{amsfonts,amssymb,amsmath}

\begin{document}

\pagestyle{empty}

\setcounter{page}{0}

\hspace{-1cm}


\begin{center}

{\LARGE{\sffamily  New path integral representation for Hubbard model: II. Spinless case}}\\[1cm]

{\large V.Kirchanov, V. Zharkov \footnote{vita@psu.ru,  Kirchanv@rambler.ru}\\[.484cm] 
\textit{ Perm State Technical university, Komsomolcky Prospect, 29a, Perm, 614600, Russia\\[.242cm]
Natural Sciences Institute of Perm State university,\\
 Genkel st.4, Perm,614990, Russia. }}
\end{center}
\vspace{2.5em}

\begin{abstract}
The Hubbard model is used to study an electronic system.
In this paper we present the new path integral representation for Hubbard
model. We have constructed the new supercoherent state for spinless electrons which appears from a set
of eigenfunctions of atomic limit of strongly correlated systems. Exact
calculation of nonlinear representation of a supergroup has been carried. This group
defines the transformation of atomic base. The general formalism we
elaborate for Hubbard model is the one widely used in the gauge field theory of the nonlinear
representation of a superconformal group.
\end{abstract}

\newpage
\pagestyle{plain}

\section{Introduction}

\noindent

The Hubbard model was originally constructed to describe
a metal-insulator transition for spin-dependent fermions
in a simple way \cite{hubbard,fradkin,fulde}.

Today this  model is still remain  the main workspace for
investigation of the strong electron's correlation.There exist many approaches to
this model for describing many electrons system: the band limit approximation
for the weak interaction between electrons and the atomic limit for electrons with the
strong coulomb repulsion. We start the series of papers in which we are intending to elaborate the new
approach to the Hubbard model.

 We will present a new path integral approarch to the strong
interaction regime. The main ingredients for us is the usage of a supercoherent state
acting upon an atomic base and the work with effective functional for an electronic
system. We will develop  the procedure of geometric quantization for strongly
correlated systems of electrons. Let us make the brief sketch of the program
firstly developed in the series of papers of one of the  authors \cite{zharkov}. One of the main
distinction of this approach to the Hubbard model is that we treat the  local
supergroup of the local space-time  as the main object of our theory. This
supergroup generates the transformation of the space-time coordinates which assigns the arguments
for any function describing this system. This representation of a supergroup in the
superspinor space has to contain a Lorents or an  SO(4) generators  equal to the even
subalgebra of the Hubbard operators. Odd Hubbard operators produce some superextension of the
Lorents group into some supergroup. This dynamic supergroup is given by 
local dynamic superfields  describing the local degrees of freedom of the
strongly correlated electrons system. We shall introduce the supercoherent
state depending on generalized angles  equal to the bosonic and fermionic
fields of the system. Parametrised by x,y,z,t space-time manifold which
determines the arguments of wave function defines us the spinor and superspinor
bundle. This superbundle is determined by the supercoherent state.

There are two kinds of the gauge fields in superspinor bundle: one sort of fields is
the composite fields equal to the quadratic combination of odd grassmann fields
and other sort are nonlinear fields determining the local coordinates frame of the
four-dimensional space-time. In general, the local superspinor bundle defines the nonlinear representation of a
superconformal group as a maximal group of 4-dimentional space-time of
interacting fermionic system.

As the first step of quantization of electronic system with strong coulomb
repulsion we perform the reformulation of Hubbard model into the atomic limit
formalism.  This approach is well known in  Mott-Hubbard insulators theory.
We want to point out that our approach includes all elements of
geometric quantisation \cite{Woodhouse}: for example possess  the algebra of 4-dimensional
rotations (Lorents) group, Cartan differential one-forms which give us the
lagrangian of system, the nonlinear representation of underlying supergroup as a
ground for the supercoherent state.

In the second paper of this series we will fulfill the exact calculation of the nonlinear
representation of spinless  supergroup generators of which appears in the  so called "tower of symmetry" in the Hubbard model \cite{zharkov}. This gives us the possibility to introduce the
supercoherent state for some deformed version of spinless algebra of the strongly correlated electron system and as a result to
obtain the effective action in a future  paper.

Let us give the brief description of the calculation method for the finding
representation of dynamic supergroup in strongly interacting models.
Our task is to find in this model such group structure which could help us
to describe the specificity of strong correlation. We take the following construction as a base:

1) we will collect space coordinates together with the time coordinate and will consider some curved
space-time as a base in which Lorents subalgebra of superconformal group act in the spinor base on dynamical fields.

2) full bases of electronic operators gives us the  supergroup which is
parametrised by  2 dynamical fermionic fields: this fields comprise the
real massless spinor.

3) the superspinor representation of a supergroup gives us the supercoherent state
described by the nonlinear function over odd grassmanian fields. This function
characterises the local properties of the strongly correlated system.

\section{Atomic description of Hubbard model}

We consider the Hubbard model:

\begin{equation}
H=-W\sum_{ij\sigma}\alpha_{\sigma,i}^{+}\alpha_{\sigma,j}+U\sum_{i,\sigma
}n_{\sigma,i}n_{-\sigma,i}+\mu \sum_{\sigma,i}n_{\sigma,i},
\end{equation}

here \ $\alpha_{\sigma,i}^{+}\alpha_{\sigma,j}$\  \ -electron creation and
annihilation operators. \ $n_{\sigma,i}$\ - electron \  \ density operator
 W, U,  \ -band width, one- site electron repulsion and chemical potential.

At first we represent the Hubbard model in atomic bases which determine the atomic limit.
This limit appear as a result of the following procedure. A zero approximation of the atomic limit is described by one-site repulsion term:
$$U\sum_{i,\sigma}n_{\sigma,i}n_{-\sigma,i}+\mu \sum_{\sigma,i}n_{\sigma
,i}.$$

This hamiltonian can be diagonalized by the following one-site atomic eigenfunction:

\begin{equation}
|0\succ;|+\succ=\alpha_{\uparrow}^{+}|0\succ;|-\succ=\alpha_{\downarrow}%
^{+}|0\succ;|2\succ=\alpha_{\uparrow}^{+}\alpha_{\downarrow}^{+}|0\succ.
\end{equation}

 This bases gives us the fundamental representation of some supergroup
in the space of dimension $(2,2)$. Point out that the states   \ $|+\succ;|-\succ;$\  \ are fermions \ but
\ $|0\succ;$\  \ $|2\succ$\  \  \ are bosons in our construction. \ We insert
later odd grassmann fields and make this fact obvious ( ie states
\ $|+\succ;|-\succ;$\ will be depend on odd order of the grassmann fields but
\ states \  \ $|0\succ;$\  \ $|2\succ$\ -on even order of the grassmann fields). All
operators in this bases will be the matrix which are determined by commutation
and anticommutation relation giving some superalgebra. Full set of the Hubbard
operators have 16 operators part of which $$(X^{0+},X^{0-},X^{+0},X^{-0}
,X^{+2},X^{-2},X^{2+},X^{2-})$$ are the fermionic operators, but other part
$$(X^{+-},X^{-+},X^{++}-X^{--},X^{02},X^{20},X^{00}-X^{22})$$--are the bosonic
operators. $X^{ij}$ -Hubbard operators contain only one non-zero element equal
1 sitting on site $(i,j)$ in the matrix representation. Point out that this set of
operators gives some bases for some superalgebra.

 We have the following representation for creation-annihilating
operators in this bases:

\begin{equation}
\alpha_{\uparrow}^{+}=X^{+0}+X^{2-}\\
\alpha_{\downarrow}^{+}=X^{-0}+X^{2+}%
\end{equation}

The Hubbard model in this representation has the form:

\begin{equation}
 H=U\sum_{i,p}X_{i}^{pp}-W\sum_{ij\alpha \beta}X_{i}^{-\alpha}X_{j}^{\beta}
\end{equation}

\section{ Supercoherent state for Hubbard model}

In constracting the supercoherent state we use the following interesting
observation in interpretation of the set of atomic operators and function for
on-site Hubbard repulsion. This observation can be formulated as the following statement: six even Hubbard
operators constitute the subalgebra isomorphic with algera of Lorentz group or algebra of four dimensional rotation group in spinor representation. Complete derivation of this statement will be obtain in subsequent paper.

 To characterise the state of the system by coherent
state we input some fields which depend on coordinates $x,y,z$ and the time $t.$
We have three component dynamic vector of the electricl field
$$
\mathbf{E}=(E^{+}(x,y,z,t),E^{-}(x,y,zt),E^{z}(x,y,z,t)),
$$
three component dynamical vector of the magnetic field
$$
\mathbf{h}=(h^{+}(x,y,z,t),h^{-}(x,y,z,t),h^{z}(x,y,z,t))
$$
and four component dynamical odd grassmann fields

$$\chi^{\ast}(x,y,z,t),\chi(x,y,z,t),$$  which are the fermionic fields giving the
components of maiorana spinor. All dynamical fields appear in supercoherent
state in the followng manner:

\begin{equation}
\label{cs}\mid G>=exp\left[
\begin{array}
[c]{cccc}%
E_{z} &\chi & \chi & E^{+}\\
\chi^{*} & h_{z} & h^{+} &\chi\\
\chi^{*} & h^{-} & -h_{z} &\chi\\
E^{-} & \chi^{*} & \chi^{*} & -E_{z}%
\end{array}
\right]  \mid0>,
\end{equation}

 Exponent here act in space of atomic eigenfunctions $(\mid
0>,\mid+>,\mid->,\mid2>)$, Function  $\mid0>$ is highest weight vector of
 representation space of supergroup which is given by exponent.

\section{ Evolution operator for electronic system}

 The transition amplitude of the evolution operator of the quantum systems is
given by the following expresion: $<Z_{f}|e^{-iH(t_{f}-t_{i})}|Z_{i}>$.
We want to obtain the expression for the effective functional using the states
$|Z>$. Time evolution of the system is given by the following operator:

$$
U(t,t_{0})=T_{ord}exp(-i\int_{t_{0}}^{t}H(\tau)d\tau);
$$

 if \ $t-t_{0}=\delta t$\  \ is small, ie $\delta t<<1$,\ then%

$$
U(t_{0}+\delta t,t_{0})=1-i\int_{t_{0}}^{t_{0}+\delta t}H(\tau)d\tau.
$$

It is follow from this expression that the symbol for evolutionary
operator has the following form:

$$
U(Z,Z^{\ast}|t_{0}+\delta t,t_{0})=exp(-i\int_{t_{0}}^{t_{0}+\delta
t}H(Z,Z^{\ast}|\tau)d\tau).
$$

 We devide time interval $[t_{0},t]$ by the number N and obtain N small
intervals for finding the expression for symbol $U(Z,Z^{\ast}|t,t_{0})$. Consider the
matrix elements of evolution operator $exp(-iH(t_{f}-t_{i}))$ between the states
$<Z_{f}|$ and $|Z_{i}>$. Factorising operator $exp(-iH(t_{f}-t_{i}))$ by inserting the identity operator
$\int d\mu(Z)|Z><Z|=1$\ we obtain the following representation:

$$
<Z_{f}|exp(-iH(t_{f}-t_{i}))|Z_{i}>=\int \prod_{k=1}^{N}d\mu(Z_{k})<Z_{f}%
|Z_{N}>
$$

$$
<Z_{N}|e^{-i\epsilon H}|Z_{N-1}>....<Z_{k-1}|e^{-i\epsilon H}|Z_{k}%
>...<Z_{1}|e^{-i\epsilon H}|Z_{i}>,
$$

 here $\epsilon=\frac{t_{f}-t_{i}}{N}.$  In first order of $\epsilon$
\ we can transform this formular and place the symbol of operator H in the exponent

\begin{equation}
\frac{<Z_{k+1}|e^{-i\epsilon H}|Z_{k}>}{<Z_{k+1}|Z_{k}>}=\frac{<Z_{k+1}%
|(1-i\epsilon H)|Z_{k}>}{<Z_{k+1}|Z_{k}>}=e^{-i\epsilon \frac{<Z_{k+1}%
|H|Z_{k}>}{<Z_{k+1}|Z_{k}>}}+O(\epsilon^{2}).
\end{equation}

 As a result we obtain the representation

\begin{equation}
<Z_{f}|e^{-iH(t_{f}-t_{i})}|Z_{i}> = \lim_{N->\infty}\int \prod_{k=1}^{N} d
\mu(Z_{k}) <Z_{k+1}|Z_{k}> e^{-i\epsilon \sum_{k=1}^{N}\frac{<Z_{k+1}|
H|Z_{k}>}{<Z_{k+1}|Z_{k}>}},
\end{equation}

here $|Z_{0}>=|Z_{i}>;<Z_{N+1}|=<Z_{f}|.$  Let define a variation of the following type $|Z>:|\delta Z_{k+1}>=|Z_{k+1}
>-|Z_{k}>.$ We have:

$$
<Z_{f}|e^{-iH(t_{f}-t_{i})}|Z_{i}> = \lim_{N->\infty}\int \prod_{k=1}^{N} [d
\mu(Z_{k}) <Z_{k}|Z_{k}>]<Z_{f}|Z_{N}>
$$

$$
exp(\sum_{k=1}^{N}(Ln(1-\frac{<Z_{k}|\delta Z_{k}>}{<Z_{k}|Z_{k}>}
)-i\epsilon \frac{<Z_{k}| H|Z_{k-1}>}{<Z_{k}|Z_{k-1}>}).
$$

 in linear-slice approximation. We take the following expresion for the time derivative: Considering the first
order in $\epsilon$\ we can take the folllowing expresion for the time derivative:

$$
\frac{d|Z>}{dt}=\frac{|\delta Z>}{\epsilon}.
$$

In first order in $\epsilon$, we obtain the final path integral representation
of the evolutionary operator in coherent state formalizm

\begin{equation}
<Z_{f}|e^{-iH(t_{f}-t_{i})}|Z_{i}>=\int_{|Z(t_{i})>=|Z_{i}>}^{|Z(t_{f}%
)>=|Z_{f}>}D(Z,Z^{\ast})e^{-iS[Z,Z^{\ast}]};
\end{equation}%
$$
S[Z,Z^{\ast}]=\int_{t_{i}}^{t_{f}}dt\int_{V}dr(\frac{<Z(r,t)|i\frac
{\partial}{\partial t}-H|Z(r,t>}{<Z(r,t|Z(r,t>}
$$
$$
-i[Ln(<Z_{f}|Z(t_{f})>)-Ln(<Z_{i}|Z(t_{i})>)]).
$$
Measure of integration is given by the following expresion:

$$
D[Z,Z^{*}] = \prod_{t_{i}<t<t_{f}} \prod_{r} d\mu[Z(r,t)^{*}
,Z(r,t)]<Z(r,t)|Z(r,t)>.
$$

 This form of path intergal representation will be the starting point
of our consideration.

\section{ Nonlinear representation of supergroup in Hubbard model}

In construction of supercoherent state we have the followng supermatrix which
we must compute analyticaly:

$$
U=\exp \left(
\begin{array}
[c]{cccc}%
E_{z} & \chi & \chi & E^{+}\\
\chi^{*} & h_{z} & h^{+} & \chi\\
\chi^{*} & h^{-} & -h_{z} & \chi\\
E^{-} & \chi^{*} & \chi^{*} & -E_{z}%
\end{array}
\right)
$$

In this expression we have the fields of different statistics: for example, set of
the fields 

$$(\chi^{*}(t,x,y,z),\chi(t,x,y,z))$$ 

are the odd grassman valued function of space-time coordinates and
describe the fermionic degree of freedom but fields 

$$(E_{z}(t,x,y,z),E^{+}
(t,x,y,z),E^{-}(t,x,y,z),h_{z}(t,x,y,z),h^{+}(t,x,y,z),h^{-}(t,x,y,z)) $$

describe the electro-magnetic degree of freedom, equal to two three component
vectors of the space-time coordinates and are bosonic. In this paper we concentrate in calculating exact representation of exponent of the supermatrix in the coherent state. We take the dynamic electrical, magnetic and grassmann fields which depend on coordinates of 4-dimensional space-time manifold on definite the coordinates and omit  $(t,x,y,z)$\ coordinates in subsiquent formulas.

Our general strategy will be to expand the supermatrix to N-order in fields. Them
we can isolate and collect certain series and get some recurrent formular for general term in infinite series. Using this formula we can sum all terms to anytical compact representation. Analytical representation of the supermatrix elements will be the final point of our work. As a starting point we have the following exponential expression for the representation of super extension of the Lorentz group in the spinor representation.

Expanding this exponent in series we can obtain first and second order in
the parameters $ b$ and $ h $. We see that the polynomial series on the grassmann numbers can be classified in grassmann order n. All the supermatrix elements can be represented as a coefficients in grassmann polynomials of order n, where n=0,1,2,3,4.

\section{ Matrix series for the nonlinear representation of supergroup}

 First of all point out that the supermatrix \ in exponent have two
submatrix: one is the odd grassmann matrix and the other is the even submatrix
containing only the fields of type $E_{i}(t,x,y,z)$ \ and $h_{i}(t,x,y,z)$ \ type.
\ As a first step we make expansion of the exponent for the even matrix.
Expanding in series this exponent we can obtain first, second order and 3,4,5
order in bosonic fields $E_{i},i=1,2,3.$ and \ $h_{i},i=1,2,3.$ For example, the series for n=0, 1,2 has the following form:

$$\left(
\begin{array}
[c]{cccc}%
1 & 0 & 0 & 0\\
0 & 1 & 0 & 0\\
0 & 0 & 1 & 0\\
0 & 0 & 0 & 1
\end{array}
\right)  +$$

$$\left(
\begin{array}
[c]{cccc}%
E_{z} & 0 & 0 & E^{+}\\
0 & h_{z} & h^{+} & 0\\
0 & h^{-} & -h_{z} & 0\\
E^{-} & 0 & 0 & -E_{z}%
\end{array}
\right)  +\frac{1}{2!}\left(
\begin{array}
[c]{cccc}%
b^{2} & 0 & 0 & 0\\
0 & h^{2} & 0 & 0\\
0 & 0 & h^{2} & 0\\
0 & 0 & 0 & b^{2}%
\end{array}
\right)  +$$

$$\frac{1}{3!}\left(
\begin{array}
[c]{cccc}%
b^{2}E_{z} & 0 & 0 & b^{2}E^{+}\\
0 & h^{2}h_{z} & h^{2}h^{+} & 0\\
0 & h^{2}h^{-} & -h^{2}h_{z} & 0\\
b^{2}E^{-} & 0 & 0 & -b^{2}E_{z}%
\end{array}
\right)  +\frac{1}{4!}\left(
\begin{array}
[c]{cccc}%
b^{4} & 0 & 0 & 0\\
0 & h^{4} & 0 & 0\\
0 & 0 & h^{4} & 0\\
0 & 0 & 0 & b^{4}%
\end{array}
\right)  +$$

$$  \frac{1}{5!}\left(
\begin{array}
[c]{cccc}%
b^{4}E_{z} & 0 & 0 & b^{4}E^{+}\\
0 & h^{4}h_{z} & h^{4}h^{+} & 0\\
0 & h^{4}h^{-} & -h^{4}h_{z} & 0\\
b^{4}E^{-} & 0 & 0 & -b^{4}E_{z}%
\end{array}
\right)  $$

 here we introduce the following abbriviation \ $b=\sqrt{E_{z}%
^{2}+E^{+}E^{-}},h=\sqrt{h_{z}^{2}+h^{+}h^{-}}$

For the odd grassmann number we \ have following expansion series of exponent. We write here two terms for the grassmann fields $(\chi^{*},\chi)$ for obtaining coefficients in higher order in $E_{i}$ \  \ and
\ $h_{i}$ .

 $$ \left(\begin{array}
[c]{cccc}%
0 & \chi & \chi & 0\\
\chi^{*} & 0 & 0 & \chi\\
\chi^{*} & 0 & 0 & \chi\\
0 & \chi^{*} & \chi^{*} & 0
\end{array}
\right)  +$$

$$\frac{1}{2!}\left(
\begin{array}
[c]{cccc}%
0 & \chi^{*} E^{+}+\chi p1 & \chi^{*} E^{+}+\chi p3 & 0\\
\chi^{*}E_{z}+\chi ( E^{-}+ h^{+}+ h_{z}) & 0 & 0 & \chi^{*} E^{+}+\chi(- E_{z}+ h^{+}+ h_{z})\\
\chi^{*} E_{z}+\chi (E^{-}+ h^{-}- h_{z}) & 0 & 0 & \chi^{*} E^{+}+\chi(- E_{z}+ h^{-}- h_{z})\\
0 & \chi^{*}p2+\chi E^{-} & \chi^{*}p4+\chi E^{-} & 0
\end{array}
\right)  $$

 For even order of the grassmann variables we have following series for the
composite bosonic fields:

$$\left(
\begin{array}
[c]{cccc}%
2 \chi \chi^{*} & 0 & 0 & 0\\
0 & 0 & 0 & 0\\
0 & 0 & 0 & 0\\
0 & 0 & 0 & 0
\end{array}
\right)$$
$$  +\frac{1}{3!}\left(
\begin{array}
[c]{cccc}%
4 E_{z} \chi \chi^{*}+(h^{+}+h^{-})\chi \chi^{*} & 0 & 0 & 0\\
0 & -2 E_{z} \chi \chi^{*} & -2 E_{z} \chi \chi^{*} & 0\\
0 & -2 E_{z} \chi \chi^{*} & -2 E_{z} \chi \chi^{*} & 0\\
0 & 0 & 0 &-4 E_{z} \chi \chi^{*}+(h^{+}+h^{-})\chi \chi^{*}
\end{array}
\right)  +$$

$$\frac{1}{4!}\left(
\begin{array}
[c]{cccc}%
2 mm \chi \chi^{*} & 0 & 0 & 4 E^{+}E_{z} \chi \chi^{*}\\
0 & -2 E_{z}(2 h_{z}+X) \chi \chi^{*} & -4 E_{z}h^{+}\chi \chi^{*} & 0\\
0 & -4 E_{z}h^{-}\chi \chi^{*} &  -2 E_{z}(-2 h_{z}+X) \chi \chi^{*} & 0\\
-4 E^{-}E_{z} \chi \chi^{8} & 0 & 0 & 2 mm \chi \chi^{*}%
\end{array}
\right)  $$

 here $ m=e^2+h^2,X=h^{+}+h^{-}, pm=h_{z}+h^{-},pp=h_{z}+h^{+},mp=h^{+}-h_{z},mm=h^{-}-h_{z}$
and $ mm=m+2 E_{z}^2+E_{z} X$
 The expansion series for third order of the grassmann variable has the
following form of first two terms
$$\frac{m}{4!} *$$
$$\left(
\begin{array}
[c]{cccc}%
0 & E^{+} \chi^{*} +p1 \chi & E^{+} \chi^{*} +p3 \chi & 0\\
E^{-} \chi +(E_{z}+pp)\chi^{*} & 0 & 0 & E^{+} \chi^{*} +(-E_{z}+pp)\chi\\
E^{-} \chi +(E_{z}+mm)\chi^{*} & 0 & 0 & E^{+} \chi^{*}+( -E_{z}+mm)\chi\\
0 & E^{-} \chi+p2 \chi^{*} & E^{-} \chi+p4 \chi^{*} & 0
\end{array}
\right)  $$
here $p1=E_{z}+pm, p2=-E_{z}+pm, p3=E_{z}+mp, p4=-E_{z}+mp $

 Expanding this series on 12 order in even and odd parameters we can
obtain the analytical representation for the supermatrix elements.

\section{ Analytical representation of supergroup}

 Having series representation for supergroup for high order in fields
we can obtain the general analytical form for the matrix elements. Our task is to
obtain exact dependence of the matrix element over component of fields:
$E_{z,}E^{+},E^{-};h_{z},h^{+},h^{-}$ but not dependence over $b$ and $h$.

For example the general form of $u_{11}$ \ supermatrix elements in orders
higher than 8 as a functions of the dynamic even and odd grassmanian fields
are given by following expressions

$$u_{11}=cosh(e)+E_{z} sinh(e)/e +$$
$$(2 f_3 +X f_2 +2 E_{z}(2 f2EE+X f1EE)+2 E_{z}^2(2 f1EE+X fEE))\chi \chi^{*}$$

here the coefficients $f_{i}, i=2,3,4$  are some series in $e$ and
$h$ variables:

$f2EE, f1EE,fEE; f2h$ have some infinite series over $e$ and $h$ variables.\

Another matrix element equal to

$$u_{21}=\chi^{*} (f_4 + (E_{z}+h^{+}+h_{z}) f_3 +  E_{z} (h^{+}+h_{z}) f_2) + \chi E^{-} (f_3+(h^{+}+h_{z})f_2)$$

and we have some function for the coefficients: $f_4,f_3,f_2$

 For $u_{31}$ we have

$$u_{31}=\chi^{*} (f_4 + (E_{z}+h^{-}- h_{z}) f_3 +  E_{z} (h^{-}- h_{z}) f_2) + \chi E^{-} (f_3+(h^{+}+h_{z})f_2)$$

 Last element of first column is

$$u_{41}=E^{-} sinh(e)/e +2 E_{z}E^{-}( 2 f1EE + X fEE)\chi \chi^{*}$$

 For coefficients in this expression we have: $ X=h^{+}+h^{-} $

 For second column we have

$$u_{12}=\chi (f_4 + (E_{z}+h^{-}+h_{z}) f_3 + E_{z} (h^{-}+h_{z}) f_2) + \chi^{*} E^{+} (f_3+(h^{-}+h_{z})f_2)$$

For $u_{22}$ we have:

$$u_{22}=cosh(h)+h_{z} sinh(h)/h + E_{z}( f2h +X h_{z} fhh +(2 h_{z}+X) f1hh)(-2\chi \chi^{*})$$

For $u_{32}$ we have:

$$u_{32}=h^{-} sinh(h)/h + E_{z}( f_2 + 2 h^{-} f1hh + h^{-}X fhh)(-2\chi \chi^{*})$$

For $u_{42}$ we have:

$$u_{42}=\chi^{*} (f_4 + (-E_{z}+h^{-}+h_{z}) f_3 -  E_{z} (h^{-}+h_{z}) f_2) + \chi E^{-} (f_3+(h^{-}+h_{z})f_2)$$

For $u_{13}$ we have:

$$u_{13}=\chi (f_4 + (E_{z}+h^{+}- h_{z}) f_3 + E_{z} (h^{+}- h_{z}) f_2) + \chi^{*} E^{+} (f_3+(h^{+}- h_{z})f_2)$$

For $u_{23}$ we have:

$$u_{23}=h^{+} sinh(h)/h + E_{z}( f_2 + 2 h^{+} f1hh + h^{+}X fhh)(-2\chi \chi^{*})$$

For $u_{33}$ we have:

$$u_{33}=cosh(h)- h_{z} sinh(h)/h + E_{z}( f2h - X h_{z} fhh +(- 2 h_{z}+X) f1hh)(-2\chi \chi^{*})$$

For $u_{43}$ we have:

$$u_{43}=\chi^{*} (f_4 + (-E_{z}+h^{+}- h_{z}) f_3 -  E_{z} (h^{+}- h_{z}) f_2) + \chi E^{-} (f_3+(h^{+}- h_{z})f_2)$$

For $u_{14}$ we have:

$$u_{14}=E^{+} sinh(e)/e +2 E_{z}E^{+}( 2 f1EE + X fEE)\chi \chi^{*}$$

For $u_{24}$ we have:

$$u_{24}=\chi (f_4 + (-E_{z}+h^{+}+h_{z}) f_3 -  E_{z} (h^{+}+h_{z}) f_2) - \chi^{*} E^{+} (f_3+(h^{+}+h_{z})f_2)$$

For $u_{34}$ we have:

$$u_{34}=\chi (f_4 + (-E_{z}+h^{-}- h_{z}) f_3 -  E_{z} (h^{-}- h_{z}) f_2) +$$
$$ \chi^{*} E^{+} (f_3+(h^{-}- h_{z})f_2)$$

For $u_{44}$ we have:

$$u_{44}=cosh(e)- E_{z} sinh(e)/e +$$
$$(2 f_3 +X f_2 - 2 E_{z}(2 f2EE+X f1EE)+2 E_{z}^2(2 f1EE+X fEE))\chi \chi^{*}$$

We see later that many series in our list are equivalent to each other. After such selection between similar ones we have only some different series.

\section{Analytical representation for series}

Collecting the terms in series expansion for $a$ and \ $b$
\ coefficients to 12 order  we obtain for example the following representation for:

$$ f_{2}=\frac{1}{3!}+\frac{b^{2}+h^{2}}{5!}+\frac
{b^{4}+h^{2}b^{2}+h^{4}}{7!}+\frac{b^{6}+h^{2}b^{4}+h^{4}b^{2}+h^{6}}%
{9!}+\frac{b^{8}+h^{2}b^{6}+h^{4}b^{4}+h^{6}b^{2}+h^{8}}{11!}+
$$
$$
\frac
{b^{10}+h^{2}b^{8}+h^{4}b^{6}+h^{6}b^{4}+h^{8}b^{2}+h^{10}}{13!}+$$

$$\frac{b^{12}+h^{2}b^{10}+h^{4}b^{8}+h^{6}b^{6}+h^{8}b^{4}+h^{10}b^{2}+h^{12}%
}{15!}+
$$
$$
\frac{b^{14}+h^{2}b^{12}+h^{4}b^{10}+h^{6}b^{8}+h^{8}b^{6}+h^{10}%
b^{4}+h^{12}b^{2}+h^{14}}{17!}+.........$$

Let us show how to sum following infinite series for $b$ \ for
example. \ We have

$$1+\frac{b^{2}}{2!}+\frac{b^{4}}{4!}+\frac{b^{6}}{6!}+\frac{b^{8}}%
{8!}+\frac{b^{10}}{10!}+......+E_{z}(1+\frac{b^{2}}{3!}+\frac{b^{4}}{5!}%
+\frac{b^{6}}{7!}+\frac{b^{8}}{9!}+\frac{b^{10}}{11!}+\frac{b^{12}}%
{13!}+........)$$

It is seen that the first series equal to $$cosh(b)=1+\frac{b^{2}%
}{2!}+\frac{b^{4}}{4!}+\frac{b^{6}}{6!}+\frac{b^{8}}{8!}+\frac{b^{10}}%
{10!}+......$$

and the second series equal to $$sinh(b)/b=1+\frac{b^{2}}{3!}+\frac{b^{4}}%
{5!}+\frac{b^{6}}{7!}+\frac{b^{8}}{9!}+\frac{b^{10}}{11!}+\frac{b^{12}}%
{13!}+........$$

For sum of two series we have the following expression  $\cosh
(b)+E_{z}\sinh(b)/b$

The main series for us is the following expansion:

$$f=\frac{1}{5!}+\frac{b^{2}+h^{2}}{7!}+\frac{b^{4}+h^{2}b^{2}+h^{4}}{9!}%
+\frac{b^{6}+h^{2}b^{4}+h^{4}b^{2}+h^{6}}{11!}+\frac{b^{8}+h^{2}b^{6}%
+h^{4}b^{4}+h^{6}b^{2}+h^{8}}{13!}+
$$
$$
\frac{b^{10}+h^{2}b^{8}+h^{4}b^{6}%
+h^{6}b^{4}+h^{8}b^{2}+h^{10}}{15!}+$$

$$\frac{b^{12}+h^{2}b^{10}+h^{4}b^{8}+h^{6}b^{6}+h^{8}b^{4}+h^{10}b^{2}+h^{12}%
}{17!}+
$$
$$
\frac{b^{14}+h^{2}b^{12}+h^{4}b^{10}+h^{6}b^{8}+h^{8}b^{6}+h^{10}%
b^{4}+h^{12}b^{2}+h^{14}}{19!}+.........$$

 Let us show how the summation of this series can be performed.  We can make the summation of subpart of hole series:

$$\frac{b^{5}}{5!}+\frac{b^{7}}{7!}+\frac{b^{9}}{9!}+\frac{b^{11}}{11!}%
+\frac{b^{13}}{13!}+...=\sinh(b)-b-\frac{b^{3}}{3!}$$

$$h^{2}(\frac{b^{7}}{7!}+\frac{b^{9}}{9!}+\frac{b^{11}}{11!}+\frac{b^{13}}%
{13!}+....)=h^{2}(\sinh(b)-b-\frac{b^{3}}{3!}-\frac{b^{5}}{5!})$$

$$h^{4}(\frac{b^{9}}{9!}+\frac{b^{11}}{11!}+\frac{b^{13}}{13!}+....)=h^{4}%
(\sinh(b)-b-\frac{b^{3}}{3!}-\frac{b^{5}}{5!}-\frac{b^{7}}{7!})$$

Having this series representation we can rewrite expression for f

$$\frac{1}{5!}+\frac{b^{2}+h^{2}}{7!}+\frac{b^{4}+h^{2}b^{2}+h^{4}}{9!}%
+\frac{b^{6}+h^{2}b^{4}+h^{4}b^{2}+h^{6}}{11!}+\frac{b^{8}+h^{2}b^{6}%
+h^{4}b^{4}+h^{6}b^{2}+h^{8}}{13!}+
$$
$$
\frac{b^{10}+h^{2}b^{8}+h^{4}b^{6}%
+h^{6}b^{4}+h^{8}b^{2}+h^{10}}{15!}+$$

$$\frac{b^{12}+h^{2}b^{10}+h^{4}b^{8}+h^{6}b^{6}+h^{8}b^{4}+h^{10}b^{2}+h^{12}%
}{17!}+....=$$

$$\frac{1}{b^{5}}(\sinh(b)-b-\frac{b^{3}}{3!})+\frac{1}{b^{7}}h^{2}%
(\sinh(b)-b-\frac{b^{3}}{3!}-\frac{b^{5}}{5!})+\frac{1}{b^{9}}h^{4}%
(\sinh(b)-b-\frac{b^{3}}{3!}-\frac{b^{5}}{5!}-\frac{b^{7}}{7!})=$$

$$\sinh(b)/b^{5}(1+\frac{h^{2}}{b^{2}}+\frac{h^{4}}{b^{4}}+...)+(-b-\frac
{b^{3}}{3!})/b^{5}(1+\frac{h^{2}}{b^{2}}+\frac{h^{4}}{b^{4}}+...)+$$

$$(-\frac{b^{5}}{5!})\frac{h^{2}}{b^{7}}(1+\frac{h^{2}}{b^{2}}+\frac{h^{4}%
}{b^{4}}+.....)+(-\frac{b^{7}}{7!})\frac{h^{4}}{b^{9}}(1+\frac{h^{2}}{b^{2}%
}+\frac{h^{4}}{b^{4}}+...)+...$$

It is seen that the series of the type \ $1+\frac{h^{2}}{b^{2}}+\frac{h^{4}%
}{b^{4}}+.....$\  \  \ describe geometric series and gives the following result: \

$$1+\frac{h^{2}}{b^{2}}+\frac{h^{4}}{b^{4}}+.....=\frac{1}{1-\frac{h^{2}}%
{b^{2}}}=\frac{b^{2}}{b^{2}-h^{2}}$$

If we insert this result we obtain the representation for:

$$\frac{1}{5!}+\frac{b^{2}+h^{2}}{7!}+\frac{b^{4}+h^{2}b^{2}+h^{4}}{9!}%
+\frac{b^{6}+h^{2}b^{4}+h^{4}b^{2}+h^{6}}{11!}+\frac{b^{8}+h^{2}b^{6}%
+h^{4}b^{4}+h^{6}b^{2}+h^{8}}{13!}+
$$
$$
\frac{b^{10}+h^{2}b^{8}+h^{4}b^{6}%
+h^{6}b^{4}+h^{8}b^{2}+h^{10}}{15!}+....$$

$$=\frac{\frac{\sinh(b)}{b^{3}}}{b^{2}-h^{2}}+(-b-\frac{b^{3}}{3!})/b^{5}%
\frac{b^{2}}{b^{2}-h^{2}}+(-\frac{b^{5}}{5!})\frac{h^{2}}{b^{7}}\frac{b^{2}%
}{b^{2}-h^{2}}+(-\frac{b^{7}}{7!})\frac{h^{4}}{b^{9}}\frac{b^{2}}{b^{2}-h^{2}%
}+...=$$

$$\frac{\frac{\sinh(b)}{b^{3}}}{b^{2}-h^{2}}-\frac{\frac{1}{b^{2}}}{b^{2}%
-h^{2}}+(-\frac{1}{3!}-\frac{h^{2}}{5!}-\frac{h^{4}}{7!}-....)\frac{1}%
{b^{2}-h^{2}}=
$$
$$
\frac{\frac{\sinh(b)}{b^{3}}}{b^{2}-h^{2}}-\frac{\frac{1}{b^{2}%
}}{b^{2}-h^{2}}+(-\sinh(h)/h^{3}+\frac{1}{h^{2}})\frac{1}{b^{2}-h^{2}}=
$$
$$
\frac{\frac{\sinh(b)}{b^{3}}-\frac{\sinh(h)}{h^{3}}}{b^{2}-h^{2}}+\frac
{1}{b^{2}h^{2}}$$

Let us consider two series: one is \ $f$ and second is \ $f_{1}$. If we multiply $f$ \  \ by coefficient $a^{5}$ and make following substitusion
\ $b->ba,h->ha$ we obtain the following series

$$a^{5}\frac{1}{5!}+a^{7}\frac{b^{2}+h^{2}}{7!}+a^{9}\frac{b^{4}+h^{2}%
b^{2}+h^{4}}{9!}+a^{11}\frac{b^{6}+h^{2}b^{4}+h^{4}b^{2}+h^{6}}{11!}+$$
$$a^{13}\frac{b^{8}+h^{2}b^{6}+h^{4}b^{4}+h^{6}b^{2}+h^{8}}{13!}+
$$
$$
a^{15}
\frac{b^{10}+h^{2}b^{8}+h^{4}b^{6}+h^{6}b^{4}+h^{8}b^{2}+h^{10}}{15!}+$$

$$a^{17}\frac{b^{12}+h^{2}b^{10}+h^{4}b^{8}+h^{6}b^{6}+h^{8}b^{4}+h^{10}%
b^{2}+h^{12}}{17!}+.....$$

It is obvious that if we take derivative we can reduce factorial in our
series for example:

$$f_{1}=\frac{\partial (a^5 f)}{\partial a},\quad f_{2}=\frac{\partial f_{1}%
}{\partial a},\quad f_{3}=\frac{\partial f_{2}}{\partial a},\quad f_{4}%
=\frac{\partial f_{3}}{\partial a}$$

here we must insert $b->a b, h->a h$ and $a$ put to 1 after differentiation.

Taking derivatives we obtain the analytical expression for $f_{i}$ :

$$f_{1}=\frac{\frac{\cosh(b)}{b^{2}}-\frac{\cosh(h)}{h^{2}}}%
{b^{2}-h^{2}};\quad f_{2}=\frac{\frac{\sinh(b)}{b}-\frac{\sinh(h)}{h}}%
{b^{2}-h^{2}};\quad f_{3}=\frac{\cosh(b)-\cosh(h)}{b^{2}-h^{2}};
$$
$$
f_{4}=\frac{b\sinh(b)-h\sinh(h)}{b^{2}-h^{2}}$$

Series for $ f_{i}$ have the following forms:

$$f_{1}=\frac{1}{4!}+\frac{b^{2}+h^{2}}{6!}+\frac{b^{4}+h^{2}%
b^{2}+h^{4}}{8!}+\frac{b^{6}+h^{2}b^{4}+h^{4}b^{2}+h^{6}}{10!}+\frac
{b^{8}+h^{2}b^{6}+h^{4}b^{4}+h^{6}b^{2}+h^{8}}{12!}+
$$
$$
\frac{b^{10}+h^{2}b^{8}+h^{4}b^{6}+h^{6}b^{4}+h^{8}b^{2}+h^{10}}{14!}+$$

$$\frac{b^{12}+h^{2}b^{10}+h^{4}b^{8}+h^{6}b^{6}+h^{8}b^{4}+h^{10}b^{2}+h^{12}%
}{16!}+
$$
$$
\frac{b^{14}+h^{2}b^{12}+h^{4}b^{10}+h^{6}b^{8}+h^{8}b^{6}+h^{10}%
b^{4}+h^{12}b^{2}+h^{14}}{17!}+....$$

$$\ f_{2}=\frac{1}{3!}+\frac{b^{2}+h^{2}}{5!}+\frac{b^{4}+h^{2}%
b^{2}+h^{4}}{7!}+\frac{b^{6}+h^{2}b^{4}+h^{4}b^{2}+h^{6}}{9!}+\frac
{b^{8}+h^{2}b^{6}+h^{4}b^{4}+h^{6}b^{2}+h^{8}}{11!}+
$$
$$
\frac{b^{10}+h^{2}%
b^{8}+h^{4}b^{6}+h^{6}b^{4}+h^{8}b^{2}+h^{10}}{13!}+$$

$$\frac{b^{12}+h^{2}b^{10}+h^{4}b^{8}+h^{6}b^{6}+h^{8}b^{4}+h^{10}b^{2}+h^{12}%
}{15!}+
$$
$$
\frac{b^{14}+h^{2}b^{12}+h^{4}b^{10}+h^{6}b^{8}+h^{8}b^{6}+h^{10}%
b^{4}+h^{12}b^{2}+h^{14}}{17!}+........$$

$$\ f_{3}=\frac{1}{2!}+\frac{b^{2}+h^{2}}{4!}+\frac{b^{4}+h^{2}b^{2}+h^{4}}%
{6!}+\frac{b^{6}+h^{2}b^{4}+h^{4}b^{2}+h^{6}}{8!}+\frac{b^{8}+h^{2}b^{6}%
+h^{4}b^{4}+h^{6}b^{2}+h^{8}}{10!}+
$$

$$
\frac{b^{10}+h^{2}b^{8}+h^{4}b^{6}%
+h^{6}b^{4}+h^{8}b^{2}+h^{10}}{12!}+$$

$$\frac{b^{12}+h^{2}b^{10}+h^{4}b^{8}+h^{6}b^{6}+h^{8}b^{4}+h^{10}b^{2}+h^{12}%
}{14!}+
$$
$$
\frac{b^{14}+h^{2}b^{12}+h^{4}b^{10}+h^{6}b^{8}+h^{8}b^{6}+h^{10}%
b^{4}+h^{12}b^{2}+h^{14}}{16!}+........$$

$$\ f_{4}=1+\frac{b^{2}+h^{2}}{2!}+\frac{b^{4}+h^{2}b^{2}+h^{4}}%
{4!}+\frac{b^{6}+h^{2}b^{4}+h^{4}b^{2}+h^{6}}{6!}+\frac{b^{8}+h^{2}b^{6}%
+h^{4}b^{4}+h^{6}b^{2}+h^{8}}{8!}+
$$
$$
\frac{b^{10}+h^{2}b^{8}+h^{4}b^{6}%
+h^{6}b^{4}+h^{8}b^{2}+h^{10}}{10!}+$$

$$\frac{b^{12}+h^{2}b^{10}+h^{4}b^{8}+h^{6}b^{6}+h^{8}b^{4}+h^{10}b^{2}+h^{12}%
}{12!}+
$$
$$
\frac{b^{14}+h^{2}b^{12}+h^{4}b^{10}+h^{6}b^{8}+h^{8}b^{6}+h^{10}%
b^{4}+h^{12}b^{2}+h^{14}}{14!}+........$$

 If we take the series for $f$ and multiply it by $b^{2}$and take
following derivative $\frac{\partial(b^{2}f)}{\partial(b^{2})}$ we obtain the
series for $fEE.$

  $$fEE=\frac{1}{5!}+\frac{2b^{2}+h^{2}}{7!}+\frac{3b^{4}%
+2h^{2}b^{2}+h^{4}}{9!}+\frac{4b^{6}+3h^{2}b^{4}+2h^{4}b^{2}+h^{6}}{11!}+
$$
$$
\frac{5b^{8}+4h^{2}b^{6}+3h^{4}b^{4}+2h^{6}b^{2}+h^{8}}{13!}+
$$
$$
\frac
{6b^{10}+5h^{2}b^{8}+4h^{4}b^{6}+3h^{6}b^{4}+2h^{8}b^{2}+h^{10}}{15!}$$

 For series $fhh$  we must make multiplication of $f$ \ on $h^{2}$
and make derivative on $h^{2}:fhh=\frac{\partial(h^{2}f)}{\partial(h^{2})}$

$$fhh=\frac{1}{5!}+\frac{2h^{2}+b^{2}}{7!}+\frac{3h^{4}+2b^{2}%
h^{2}+b^{4}}{9!}+\frac{4h^{6}+3b^{2}h^{4}+2b^{4}h^{2}+b^{6}}{11!}+
$$
$$
\frac{5h^{8}+4b^{2}h^{6}+3b^{4}h^{4}+2b^{6}h^{2}+b^{8}}{13!}+
$$
$$
\frac{6h^{10}+5b^{2}h^{8}+4b^{4}h^{6}+3b^{6}h^{4}+2b^{8}h^{2}+b^{10}}{15!}$$

$$f_{2}EE=\frac{1}{3!}+\frac{2b^{2}+h^{2}}{5!}+\frac{3b^{4}%
+2h^{2}b^{2}+h^{4}}{7!}+\frac{4b^{6}+3h^{2}b^{4}+2h^{4}b^{2}+h^{6}}{9!}+
$$
$$
\frac{5b^{8}+4h^{2}b^{6}+3h^{4}b^{4}+2h^{6}b^{2}+h^{8}}{11!}+
$$
$$
\frac
{6b^{10}+5h^{2}b^{8}+4h^{4}b^{6}+3h^{6}b^{4}+2h^{8}b^{2}+h^{10}}{13!}$$

 To evaluate $f1EE,f2EE,f1hh$ we multiply $fEE$ by $a^{5}$
and make the following substitution  $b-> ba, h->ha.$ After calculation we fix $a=1.$

It is seen that expression for $f1EE,f2EE,f1hh$ are the following:
 $$f_{1}EE=(\frac{\partial fEE}{\partial a})_{a=1}$$

$$f1EE=\frac{1}{4!}+\frac{2b^{2}+h^{2}}{6!}+\frac{3b^{4}+2h^{2}%
b^{2}+h^{4}}{8!}+\frac{4b^{6}+3h^{2}b^{4}+2h^{4}b^{2}+h^{6}}{10!}+
$$
$$
\frac
{5b^{8}+4h^{2}b^{6}+3h^{4}b^{4}+2h^{6}b^{2}+h^{8}}{12!}+
$$
$$
\frac{6b^{10}%
+5h^{2}b^{8}+4h^{4}b^{6}+3h^{6}b^{4}+2h^{8}b^{2}+h^{10}}{14!}+........ $$

Repeating all operation we can obtain for \ $f_{2}EE=(\frac
{\partial^{2}fEE}{\partial^{2}a})_{a=1}$

$$f_{2}EE=\frac{1}{3!}+\frac{2b^{2}+h^{2}}{5!}+\frac{3b^{4}+2h^{2}%
b^{2}+h^{4}}{7!}+\frac{4b^{6}+3h^{2}b^{4}+2h^{4}b^{2}+h^{6}}{9!}+
$$
$$
\frac{5b^{8}+4h^{2}b^{6}+3h^{4}b^{4}+2h^{6}b^{2}+h^{8}}{11!}+
$$
$$
\frac{6b^{10}%
+5h^{2}b^{8}+4h^{4}b^{6}+3h^{6}b^{4}+2h^{8}b^{2}+h^{10}}{13!}$$

and for  $ f1hh=(\frac{\partial fhh}{\partial a})_{a=1}$  and for  $ f2hh=(\frac{\partial f1hh}{\partial a})_{a=1}$

$$f1hh=\frac{1}{4!}+\frac{2h^{2}+b^{2}}{6!}+\frac{3h^{4}+2b^{2}%
h^{2}+b^{4}}{8!}+\frac{4h^{6}+3b^{2}h^{4}+2b^{4}h^{2}+b^{6}}{10!}+
$$
$$
\frac{5h^{8}+4b^{2}h^{6}+3b^{4}h^{4}+2b^{6}h^{2}+b^{8}}{12!}+
$$
$$
\frac{6h^{10}+5b^{2}h^{8}+4b^{4}h^{6}+3b^{6}h^{4}+2b^{8}h^{2}+b^{10}}{14!}+.......$$

$$f2h=\frac{1}{3!}+\frac{3h^{2}+b^{2}}{5!}+\frac{5h^{4}+3b^{2}%
h^{2}+b^{4}}{7!}+\frac{7h^{6}+5b^{2}h^{4}+3b^{4}h^{2}+b^{6}}{9!}+
$$
$$
\frac{9h^{8}+7b^{2}h^{6}+5b^{4}h^{4}+3b^{6}h^{2}+b^{8}}{11!}+
$$
$$
\frac{11h^{10}+9b^{2}h^{8}+7b^{4}h^{6}+5b^{6}h^{4}+3b^{8}h^{2}+b^{10}}{13!}+.......$$

In analytical form we have for $ f2h= 2 f2hh-f_2 $

 Series for $DEDEf$ \ equal to derivative of $fEE$ on $b^{2}$. It is
seen if we compare series for $fEE$ and $DEDEf$: \  \  \ $DEDEf=\frac{\partial
fEE}{\partial(b^{2})}$

$$DEDEf=\frac{2}{7!}+\frac{6b^{2}+2h^{2}}{9!}+\frac{12b^{4}%
+6h^{2}b^{2}+2h^{4}}{11!}+\frac{20b^{6}+12h^{2}b^{4}+6h^{4}b^{2}+2h^{6}}%
{13!}+
$$
$$
\frac{30b^{8}+20h^{2}b^{6}+12h^{4}b^{4}+6h^{6}b^{2}+2h^{8}}%
{15!}+........$$

We can obtain the following representation for $DEDEf1:$ \  \  \ $DEDEf1=\frac
{\partial fhh}{\partial(b^{2})}$

$$ DEDEf1=\frac{2}{6!}+\frac{6b^{2}+2h^{2}}{8!}+\frac{12b^{4}%
+6h^{2}b^{2}+2h^{4}}{10!}+\frac{20b^{6}+12h^{2}b^{4}+6h^{4}b^{2}+2h^{6}}%
{12!}+
$$
$$
\frac{30b^{8}+20h^{2}b^{6}+12h^{4}b^{4}+6h^{6}b^{2}+2h^{8}}%
{14!}+........$$

Comparing series for $DEDhf2$ and series for $f_{2}$ we can obtain
$DEDhf2=\frac{\partial^{2}f_{2}}{\partial(b^{2})\partial(h^{2})}$

$$DEDhf2=\frac{1}{7!}+\frac{2b^{2}+2h^{2}}{9!}+\frac{3b^{4}%
+4h^{2}b^{2}+3h^{4}}{11!}+\frac{4b^{6}+6h^{2}b^{4}+6h^{4}b^{2}+4h^{6}}{13!}+
$$
$$
\frac{5b^{8}+8h^{2}b^{6}+9h^{4}b^{4}+8h^{6}b^{2}+5h^{8}}{15!}+$$

$$\frac{6b^{10}+10h^{2}b^{8}+12h^{4}b^{6}+12h^{6}b^{4}+10h^{8}b^{2}+6h^{10}%
}{17!}+.......$$

and for $DEDhf3$ we can obtain the following representation:
$DEDhf3=\frac{\partial^{2}f_{3}}{\partial(b^{2})\partial(h^{2})}$

$$
 DEDhf3=\frac{1}{6!}+\frac{2b^{2}+2h^{2}}{8!}+\frac{3b^{4}
+4h^{2}b^{2}+3h^{4}}{10!}+\frac{4b^{6}+6h^{2}b^{4}+6h^{4}b^{2}+4h^{6}}
{12!}+
$$
$$
\frac{5b^{8}+8h^{2}b^{6}+9h^{4}b^{4}+8h^{6}b^{2}+5h^{8}}{14!}+$$

$$
\frac{6b^{10}+10h^{2}b^{8}+12h^{4}b^{6}+12h^{6}b^{4}+10h^{8}b^{2}+6h^{10}}{16!}+.......$$

Summing our calculation let's write all analytical formula for the function in the matrix elements:

$$fEE=\frac{2\sinh(h)b^{3}+h\left(  b^{2}-h^{2}\right)  \cosh
(b)b+\left(  h^{3}-3b^{2}h\right)  \sinh(b)}{2b^{3}h\left(  b^{2}%
-h^{2}\right)  ^{2}};
$$
$$
f1EE=\frac{-2b\cosh(b)+2b\cosh(h)+\left(b^{2}-h^{2}\right)  \sinh(b)}{2b\left(  b^{2}-h^{2}\right)  ^{2}};$$

$$f2EE=\frac{b\left(  b^{2}-h^{2}\right)  \cosh(b)-\left(  b^{2}%
+h^{2}\right)  \sinh(b)+2bh\sinh(h)}{2b\left(  b^{2}-h^{2}\right)  ^{2}}$$

$$f2=\frac{h\sinh(b)-b\sinh(h)}{b^{3}h-bh^{3}}; \quad f3=\frac
{\cosh(b)-\cosh(h)}{b^{2}-h^{2}};
$$
$$
f4=\frac{b\sinh(b)-h\sinh
(h)}{b^{2}-h^{2}};\qquad f=\frac{\frac{\sinh(b)}{b^{3}}-\frac
{\sinh(h)}{h^{3}}}{b^{2}-h^{2}}$$

\section{ Conclusion}

 We have calculated the exact representation of the supergroup as well as the
supercoherent state in the Hubbard model. These constructions naturally appear in the
strongly correlated electronic systems in the case of introducing atomic bases in limit
of large on-site Hubbard repulsion. The dynamical supergroup which operates in a
local superbundle determined by any on-site eigenfunction gives us the wave function
in the form of a superspinor. This superspinor describes a local supercoordinates
frame in the curved supermanifold. The operator spinor part acting in tangent and
cotangent bandles of this supermanifold in supergroup can be reformulated in the
terms of the atomic Hubbard operators. Next step lies in the calculation of
effective functional of Hubbard model.

\end{document}